\documentclass[12pt]{article}
\usepackage{epsfig}

\def\beq{\begin{equation}}
\def\eeq#1{\label{#1}\end{equation}}
\def\eeqn{\end{equation}}

\def\beqa{\begin{eqnarray}}
\def\eeqa#1{\label{#1}\end{eqnarray}}
\def\eeqan{\end{eqnarray}}

\let\bar=\overbar

\def\Dslash{\not{\hbox{\kern-4pt $D$}}}
\def\dslash{\not{\hbox{\kern-2pt $\del$}}}

\def\msb{{\bar{\ssstyle M \kern -1pt S}}}

\usepackage{fancyhdr,graphicx}
\def\Title#1{\begin{center} {\Large {\bf #1} } \end{center}}

\begin{document}

\Title{Cosmic Matter under Extreme Conditions:\\
CSQCD II Summary}

\bigskip\bigskip

%+\addcontentsline{toc}{chapter}{{\it I. Med}}
%+\label{MedInseneStart}

\begin{raggedright}

{\it Jochen Wambach\index{Med, I.}\\
Institut f\"ur Kernphysik\\
Technical University Darmstadt\\
Schlossgartenstr. 9\\
D-64289 Darmstadt\\
Germany\\
{\tt Email: wambach@physik.tu-darmstadt.de}}
\bigskip\bigskip
\end{raggedright}

\abstract
After the first meeting in Copenhagen in 2001 QSQCD II is the second workshop in this series dealing with cosmic matter at very high density and its astrophysical implications. The aim is to bring together reseachers in the physics of compact stars, both theoretical and observational. Consequently a broad range of topics was presented, reviewing extremely energetic cosmological events and their relation to the high-density equation of state of strong-interaction matter. This summary elucidates recent progress in the field, as presented by the participants,  and comments on pertinent questions for future developments.

\section{Introduction}

Compact stars such as white dwarfs, neutron stars or black holes are the densest objects in the Universe.
They are the final product of stellar evolution.  Which object is formed depends on the mass of the progenitor star. Since at least half of the stars in our galaxy are bound in pairs one is often dealing with binary systems with interesting evolution histories such as merger events or complex accretion phenomena. 

Neutron- hyprid- or quark stars are the most compact objects without an event horizon. Their observational signatures, composition and evolution and the relation to states of strong-interaction matter was the major focus of this workshop.  For the cosmological events in question one deals with matter at interparticle spacings of the order of femtometers and temperatures ranging from $10^6$ K to $10^{11}$ K. The immediate consequence is that, in constrast to ordinary materials whose thermodynamic states are governed by the electromagnetic force,  now all four fundamental forces are involved in a major way. Their interplay leads to rich and complex phenomena not found under terrestrial conditions. The description of these phenomena encompasses many fields of physics such as nuclear and particle physics, relativistic fluid dynamics and general relativity, condensed matter physics and the properties of ultracold atomic gases, plasma physics, turbulence etc. in a truely interdisciplinary way.\\

\noindent
More spefically, questions that have also been adressed during the meeting include
\begin{itemize}
\item what are the observational and evolutionary manifestations of compact stars? 
\item how does the quark-hadron transition influence compact star composition and evolution? 
\item what is the equation of state (EoS) of extremely heated and dense matter?
\item how does compact star matter couple to the electro-weak sector of the standard model?
\item what is the influence of strong magnetic fields as encountered in magnetars?
\item is quark matter the absolute ground state and what would be its astrophysical consequences?
\end{itemize}

The workshop was organized in three sections. The first part was devoted to observations of cosmic events and how they relate to compact objects in the universe. The second part dealt with theoretical and observational issues of the transition from neutron stars to quark stars. The third part focused mainly on the EoS of high-density matter and the quark-hadron phase transition. 

\section{Observational manifestations of compact stars}

Since the discovery of neutron stars as radio pulsars in 1967 by Bell and Hewish, observers have collected an impressive amount of data on compact objects in the galaxy. A traditional method is the timing behavior of pulsars with radio telescopes. As discussed by G. Hobbs, timing observations have come to a remarkable state that now makes it possible to probe neutron star properties such as maximum masses, moments of inertia etc. in great detail.  Because of mass-shedding, the rotation period severely constraints the mass-radius relation of a neutron star and thus directly relates to the EoS of high density matter (C. Zhang). The most rapidly rotating pulsar is PSR J1748-2446ad with rotation period of 1.4 ms, observed in 2004. At present its puts the tightest constraint on the mass-radius relation \cite{LatPrak}. However, recently a 0.89 ms X-ray burst oscillation has been reported in the X-ray transient XTE J1739-285. This is rather hard to acommodate with ordinary equations of state and may hint to existence of quark stars (C. Zhang).
   
Rotation frequencies and spin-down rates for close to a thousand pulsars have been measured that tell about the evolutionary history of neutron stars in isolation or in binaries. The timing of glitches - sudden speed ups of the rotation period relaxing back in days to years - gives information on the composition and dynamics of the outer crust of the star through post-glitch relaxation. The spin-down rate of the youngest pulsars is dominated by these relaxation as can be infered from the timing noise (G. Hobbs).

As mentioned above, many compact stars come in binaries. Of special importance are X-ray binaries with luminosities of $10^{35}-10^{39}$ erg/s in the X-ray band (M. Gilfanov, X. Li). About 200 bright binary X-ray sources are known in the Milky Way. In these systems matter is accreted onto the compact object, either from a high-mass companion (HMXB) in which the companion is an O- or B star or a blue supergiant,  or from an object less massive than the accreter (LMXB) such as a main sequence star, an evolved red giant or a white dwarf. HXMB's evolve over time scales of 10-100 Myrs while LMXB's accrete for 1-10 Gyrs, a typical lifetime of the host galaxy. Luminosity functions and population estimates for both classes were discussed. 

Gamma-ray bursts (GRBs) are flashes of gamma rays in distant galaxies associated with extremely energetic explosions with isotropic energy outputs of up to $10^{55}$ erg (B. Zhang, Y.F. Huang). They are the most luminous electromagnetic events occurring in the universe and are extremely rare (a few per galaxy per million years). Bursts can last from milliseconds to nearly an hour, although a typical burst duration is a few seconds. The initial burst is usually followed by a longer-lived 'afterglow' emitting at longer wavelengths. In almost half of the cases a bump ('X-ray flair') appears on top of the smooth afterglow light curve. Most observed GRBs are believed to be a narrow beam of intense radiation released during a supernova event, as a rapidly rotating, ultra-high-mass star collapses to form a black hole ('long bursts'). A subclass of GRBs ('short' bursts) appears to originate from a different process, possibly the merger of binary neutron stars or a neutron star and a black hole. The means by which gamma-ray bursts convert energy into radiation is not fully understood. Of special interest to the workshop was the question of what the 'central engine' is that powers GRBs. As discussed by B. Zhang this is not yet fully understood but observational constraints from the variability of the light curves on its size can be given. From the observation of sub-millisecond variability time scales one concludes that the central engine must be very compact (less than $10^7$ cm) pointing a black hole or a compact star as central engine candidates. The flairs in the afterglow light curves are interpreted as late activities of the central engine. Generally three types of central engines are presently favored: black holes with thick accretion disks (black-hole torus systems), millisecond magentars or strange quark stars. The latter will be discussed in more detail later. Another possibility for the GRB mechanism, consistent with the observational constraints, is neutron star kicks driven by electromagnetic radiation (Y. F. Huang).  

\section{The properties of compact stars}

With central energy densities exceeding those of saturated nuclear matter by several times, compact stars are extraordinary laboratories for dense matter physics and thus directly relate to the phase diagram of matter under extreme conditions (R. Xu). The ultimate fate of matter under extreme conditions in temperature and density has intrigued physicists for a long time.  An early phase diagram from 1953 can be found in E. Fermi's book 'Notes on Thermodynamics and Statistics'\cite{Fermi}. As a function of temperature and pressure it features the then known 'elementary particles' namely protons, neutrons and electrons\footnote{Although particles such as the muon, the pion and the kaon had been discovered by 1953, it was not yet clear how they feature in the phase diagram.}. The situation changed dramatically in the 1960's when a plethora of new particles was discovered in accelerators. It led to the development of the quark model by Gell-Mann, Nishijima, Ne'eman and Zweig and eventually to Quantum-Chromodynamics (QCD) {FGL}. The fact that hadrons are composites opened the possilility for a 'quark-hadron' transition in which the (confined) quarks and gluons are liberated to form quark-gluon matter. The transition not only happens at high temperatures but also at high densities and low temperatues, the realm of supernova matter and compact stars. A 'modern' view of the phase diagram and its salient features was presented by D. Blaschke. 

One of the central questions of this meeting was, how to assess the strong-interaction phase diagram theoretically?  Since QCD is the fundamental theory of the strong interaction, why does one not use it directly to compute the phase diagram from first principles?  Ab-initio calculations of the equilibrium thermodynamics start from the QCD Lagrangian. The grand-canonical partition function is written down in euclidean space-time as a path integral which is then discretized on a four-dimensional lattice. The partition function can be evaluated stochastically by Monte-Carlo methods at vanishing chemical potential, $\mu$ and works very well for the EoS in this regime. As was pointed out by K. Splittorf, serious problems arise at finite $\mu$.  Using chiral pertubation theory - an expansion of the QCD action in the pion mass $m_\pi$ -  the conclusion is that the sign problem is tractable for $\mu<m_\pi/2$  but becomes severe when $\mu>m_\pi$. However, this is exactly the region of interest for the transition from nuclear to quark matter at low temperatures and hence for the EoS for compact star matter! Whether or not the sign problem can be overcome in the future remains to be seen. 
 
At high energies the QCD coupling becomes small and pertubation theory is applicable. This can be exploited to make rigorous statements about the state of three-flavor cold quark matter (T. Sch\"afer) \cite{ASRS}. At very large $\mu$ one can set up an effective theory in which quarks behave like non-relativistic quasi-particles and -holes near the Fermi surface. The interaction is mediated by screened longitudinal and Landau-damped transverse gluons and leads to the superconducting color-flavor-locked (CFL) phase with a paring gap $\Delta$. Chiral symmetry is spontaneously broken and Goldstone bosons arise, much like in the QCD vacuum. In addition all eight gluons acquire a mass.  From an observational point of view an interesting question is, how CFL matter and other superconduting phase responds to very strong magnetic fields ($10^{15}-10^{16}$ G) (V. de la Incera, E. Ferrer, B. Niebergal) and what the implications for magnetars are. In particular, strong magnetic fields may lead to inhomogeneous gluon condensation and the appearence of vortex structures and might be able to explain problems with standard magnetar models. Another question is the origin of very strong magnetic fields? One possiblity is the spontaneous magnetization of quark matter (T. Tatsumi).  

As the density is lowered the fact that the strange quark mass $m_s$ is significantly larger than the up and down masses comes into play. As a consequence the Fermi surface of the strange quarks is no longer equal to that of the up and down quarks which starts to inhibit the pairing between light- and strange quarks. The question then is, how the matter responds in detail to the mass imbalance. This can be systematically studied in a chiral effective theory for Golstone bosons and massive composite fermions. Based on this theory the phase ordering can be predicted by systematically studying the response to the imbalance $\delta\mu$ of the Fermi surfaces of the up-, down- and strange quarks. While for $\delta\mu\ll \Delta$ quark matter is in the CFL phase, the regime $\delta\mu\sim \Delta$ is quite complicated, featuring $s$- and $p$-wave meson condensation. 

For neutron star evolution the transport properties of quark matter - if it indeed exists in the inner core or is absolutely stable - are of relevance. During the internal motion of the star, its energy can be dissipated viscously or through heat transport. The bulk and shear viscosities, for instance, play an important role in the damping of r-mode oscillations that couple to gravitational waves (P. Jaikamur). In addition, neutron stars can loose energy efficiently through neutrino emission and one would like to know the emissivities and the neutrino mean free path (T. Schaefer, Q. Wang). The transport coefficients are governed by the quasiparticles as the elementary excitations of the matter and can be calculated in kinetic theory as scatterings between these quasiparticles. Leading-order results for both unpaired quark matter and the CFL phase are now available. While in the former the transport properties are determined by the electric screening mass of the gluons, in the CFL phase Goldstone mode scattering and decay play the decisive role (T. Sch\"afer).

The question remains, whether the ultra-high density limit of QCD is of relevance for the actual central densities encountered in compact stars. An alternative to lattice simulations of the EoS at finite density is the continuum Dyson-Schwinger approach in which the quark- and gluon propagators are calculated in a particular gauge with appropriate truncations of the hierarchy of coupled Green's functions. This approach does not suffer from the sign problem and aspects of the phase diagram such as chiral symmetry restoration and the deconfinement transition can be studied both at finite $T$ and finite $\mu$ by modelling the quark-gluon vertex or adjusting it to lattice QCD results (Y.-X. Liu). In addition color superconducting properties can be calculated at densities relevant to compact stars. A first attempt to infer the quark momentum distribution, the bag pressure and the quark matter EoS from QCD Dyson-Schwinger equations was presented by T. Kl\"ahn. 

The most detailed calculations of the strong-interaction phase diagram have been obtained within chiral quark models , in which the gluon degrees of freedom are frozen in terms of effective quark-quark of quark-meson couplings (Y.-X. Liu, D. Blaschke) These are then adjusted to vacuum properties like meson masses and the weak pseudoscalar decay constants. To account for aspects of deconfinement the quarks are coupled to a background temporal Polyakov field that statistically supresses the quark degrees of freedom in the confined phase. The calculated thermodynamics is in very good agreement with recent lattice QCD results at vanishing $\mu$. These models can thus be used with some confidence to study the BCS-BEC crossover and pion condensation at finite isospin chemical potential (P. Zhuang, C. Mu), various low-density superconducting phases, such the 2SC phase in which only up- and down quarks pair, as well as gapless phases. As these calculations are mostly performed at the mean-field level, there is still a long way to go to make contact to cold nuclear and neutron matter, which is important for neutron star crustal and outer liquid core properties (L. W. Chen) . A proper treatment at least requires the inclusion of three quark correlations, which constitues an in-medium Fadeev problem (D. Blaschke).

\section{From neutron stars to quark stars}

It is natural to ask what the observable consequences for quark cores in ordinary neutron stars or the existence of absolutely stable quark stars are? 

As discussed by T. Fischer, one possible signal is a second neutrino burst in core collapse supernovae. When the iron core of a massive progenitor star ($M> 8-9 M_\odot$) collapses gravitationally, an outward propagating shock wave is formed leaving behing a hot proto-neutron star with a radius of about 100 km. As the proto-neutron star continues to accrete infalling matter and cools, conditions in density and temperature are reached that might induce a quark-hadron transition, leading to the formation of a second shock front. This second shock might be energetic enough to explain supernova explosions even in spherical geometry. When the secondary shock hits the neutrino sphere, a second neutrino burst  - mostly composed of electron antineutrinos - is created and should be observable in next-generation neutrino detectors. The explosion simulations, presented by Fischer, were done with rather simplified equations of state for the quark-hadron transition, which might not be realistic. Since the second $\nu$ burst is a clear prediction, one can turn the argument around and use its presence or absence in the supernova neutrino spectrum as an observational constraint on the high-density EoS.

It has been proposed by Bodmer, Terazawa and Witten long ago, that $^{56}Fe$ (or nuclear matter in bulk) may not be the lowest-energy state of matter at zero pressure. Due an energy gain by converting u- or d-quarks into strange quarks, thus reducing the Pauli repulsion, there may exist a lower-energy selfbound state of three flavor quark matter, usually called 'strange quark matter' (SQM) \cite{Mad}. According to what has been discussed earlier it is most likely a color superconductor in the CFL phase.  SQM can exist in droplets that may vary in size from femtometers (strangelets) to kilometers (strange quark stars). Thus, nuclear matter is actually metastable and, given enough time, converts into SQM. In this process the extra binding energy of about 50 MeV/baryon is released which, in the astophysical context, could lead to the occurrence of quark novae (R. Ouyed). A quark nova is a hypothetical type of supernova that would be initiated if a young proto-neutron star or a spinning-down neutron star were to spontaneously convert into a self-bound quark star. The amounts of energy released may explain the most energetic supernova explosions such as SN2006gy with unsusally high energy fractions in electromagnetic radiation. This makes quark stars candidates for the central engine of GRB's (J. Staff). It also has implication for r-process nucleosynthesis in supernova explosions (R. Ouyed). J. Horvarth discussed details of the neutron star to quark star conversion mechanisms. As he pointed out, the energy release happens combustively. Due to instabilites, the burning front most likely evolves in a turbulent manner much like the burning processes in white dwarfs, that lead to Type Ia supervovae. The time scales are of the order of milliseconds and may lead to a turbulent deflagration or a detonation. Those could revive the stalled front of the prompt shock in a type II supernova to lead to a successful explosion. Other energy transfer mechanisms could include shock heating through SQM neutrinos or photons created at the strange star surface (K.S. Cheng).

As a consequence of di-quark or multi-quark correlations cold quark matter may be in a solid state giving rise to elastic properties (R. Xu, W. Fu). Thus, non-radial speriodal and toriodal modes can be excited. While such modes also exist in the solid crust of an ordinary neutron star, detailed analyses of quasi-periodic oscillation (GPO) timing signals of soft gamma repeaters like SGR 1806-20 and other magnetars  may hint towards solid strange quark stars or at least to contributions from crystalline superconducting phases in the quark core of an ordinary neutron star (W. Fu).

\section{Conclusions}

Since the first CSQCD meeting eight years ago significant progress has been made in the physics of compact star physics both observationally and in the development of theoretical tools, as we have witnessed during this meeting. With the advent of new space based telescopes such as the Fermi telescoe, launched in 2008, or Constellation-X in the future as well as with further improvements in the theoretical methods to assess the high-density EoS of strong-interaction matter, we can look forward to a bright future of this field. 

Given the broad range of topics covered during the workshop, it is difficult to do justice to all aspects discussed and I apologized to all those who do not feel adequately represented in this summary. Let me also take the opportunity to thank once more our hosts at the Kavli Institute for Astronomy and Astrophysics in Beijing for providing an excellent environment as well as the organizing committee for putting together a fine program.


\bigskip


\begin{thebibliography}{99}

%%
%%  bibliographic items can be constructed using the LaTeX format in SPIRES:
%%    see    http://www.slac.stanford.edu/spires/hep/latex.html
%%  SPIRES will also supply the CITATION line information; please include it.
%%

\bibitem{LatPrak}
    J. Lattimer and M. Prakash, Phys. Rept. {\bf 442}, 109 (2007).

\bibitem{Fermi}
E. Fermi, {\it Notes on Thermodynamics and Statistics}, University of Chicago Press, 1966. 

\bibitem{FGL} H. Fritzsch, M. Gell-Mann and H. Leutwyler, Phys. Lett {\bf 47B}, 365 (1973).

\bibitem{ASRS} M. Alford, A. Schmitt, K. Rajagopal and T. Sch\"afer, Rev. Mod. Phys. {\bf 80}, 1455 (2008).

\bibitem{Mad} J. Madsen, Lect. Notes Phys. {\bf 516}, 162 (1999).

\end{thebibliography}
\end{document}